\documentclass[letterpaper]{article} 
\usepackage[draft]{aaai2026}  
\usepackage{times}  
\usepackage{helvet}  
\usepackage{courier}  
\usepackage[hyphens]{url}  
\usepackage{graphicx} 
\usepackage{natbib}  
\usepackage{caption} 
\usepackage{amssymb}
\usepackage{amsmath}
\usepackage{array}
\usepackage{subcaption}
\usepackage{algorithm}
\usepackage{algpseudocode}
\usepackage{placeins}
\usepackage{xcolor}

\usepackage{newfloat}
\usepackage{listings}
\DeclareCaptionStyle{ruled}{labelfont=normalfont,labelsep=colon,strut=off} 
\floatstyle{ruled}
\newfloat{listing}{tb}{lst}{}
\floatname{listing}{Listing}
\usepackage{xspace}

\title{Hybrid restricted master problem for Boolean matrix factorisation}
\author {
    Ellen Visscher\textsuperscript{\rm 1},
    Michael Forbes\textsuperscript{\rm 2},
    Christopher Yau\textsuperscript{\rm 1,3},
}
\affiliations {
    \textsuperscript{\rm 1}University of Oxford\\
    \textsuperscript{\rm 2}University of Queensland\\
    \textsuperscript{\rm 3}Health Data Research UK\\
    ellen.visscher@bdi.ox.ac.uk, m.forbes@uq.edu.au, christopher.yau@wrh.ox.ac.uk}
    
\newcommand{\bfeps}{\text{\Large $\boldsymbol{\epsilon}$}}
\newcommand{\eps}{\text{\Large $\epsilon$}}

\newcommand{\bfact}{\texttt{bfact}\xspace}

\makeatletter
\def\clearallfloats{%
  \global\let\@currbox\@empty
  \global\let\@deferlist\@empty
  \global\let\@freelist\@empty
  \global\let\@dblfloat\@empty
  \global\let\@fls\@empty
  \global\let\@fpsbox\@empty
}
\makeatother
\begin{document}

\maketitle

\begin{abstract}
We present \bfact, a Python package for performing accurate low-rank Boolean matrix factorisation (BMF). \bfact uses a hybrid combinatorial optimisation approach based on \emph{a priori} candidate factors generated from clustering algorithms. It selects the best disjoint factors before performing either a second combinatorial or heuristic algorithm to recover the BMF. We show that \bfact does particularly well at estimating the true rank of matrices in simulated settings. In real benchmarks, using a collation of single-cell RNA-sequencing datasets from the Human Lung Cell Atlas, we show that \bfact achieves strong signal recovery, with a much lower rank.

\end{abstract}

%
\begin{links}
    \link{Package Code}{https://github.com/e-vissch/bfact-core}
    \link{Analysis Code}{https://zenodo.org/records/17589218}
\end{links}

\section{Introduction}

Matrix factorisation techniques are frequently used in machine learning to identify latent structure in data sets by decomposing them into lower-dimensional representations that capture common underlying patterns or concepts. Popular matrix factorisation techniques include Singular Value Decomposition (SVD), Principal Component Analysis (PCA), and Nonnegative Matrix Factorisation (NMF). SVD and PCA require orthogonal factors, while NMF constrains the target matrix and the factors to be nonnegative. In the special case where the input is a binary matrix, Boolean matrix factorisation (BMF) seeks to produce two low-rank Boolean factor matrices whose Boolean product is close to the original input. With binary outputs, BMF preserves a form of interpretability with one factor matrix composed of a limited set of binary basis vectors capturing combinations of frequently co-occurring raw features, while the other associates input samples to one or more basis vectors. This has seen its use in many data mining applications in market basket analysis, bioinformatics and recommender systems.

\subsection{Motivation}

In this work, we are particularly motivated by the use of BMF for applications in biology and specifically, that of single-cell RNA sequencing analysis (scRNAseq). Single-cell technologies now enable biologists to routinely screen hundreds of thousands of cells for the activity (or \emph{expression}) of tens of thousands of genes. The data matrices are therefore significant (e.g $\sim100$k $\times 15$k) and a common task is to perform a form of dimensionality reduction to interpret the data by projecting onto low-dimensional representations using techniques such as principal components analysis (PCA), non-negative matrix factorisation (NNMF) \cite{qi_clustering_2020, wang_unsupervised_2022, tsuyuzaki_benchmarking_2020}, visualisation methods (t-SNE, UMAP) and deep learning approaches \cite{wu_tools_2020, wang_comparison_2022}. However, single-cell data can often be approximated as binary due to the relatively sparse number of sequencing reads per cell, making the use of BMF amenable \cite{rukat_bayesian_2017, liang_bem_2020}.
Further, the direct interpretation of factor feature (gene) sets is useful for downstream analysis. 

\subsection{Theory}

We first introduce the formal mathematical background.
We wish to decompose $\mathbf{X} \in \{0, 1\}^{M \times N}$ into two other low-rank binary matrices, $\mathbf{L} \in \{0, 1\}^{M \times K}$, $\mathbf{R} \in \{0, 1\}^{K \times N}$ such that the original matrix can be recapitulated using Boolean logic according to $X_{ij} = \bigvee_{k=1}^K L_{ik} \land R_{kj}$. Here, $(M, N)$ corresponds to the number of observations and features, respectively and typically $K \ll N$. In practice, observations are corrupted by noise and it is common to assume that the observations $\mathbf{Y} \in \{0, 1\}^{M \times N} = \mathbf{X} + \bfeps$ where $\bfeps$ is an additive noise matrix such that $\epsilon_{ij} \in \{-1,0,1\}$. The objective is to recover the signal matrix $\mathbf{X}$ through finding the lowest rank $\mathbf{\hat{L}}$ and $\mathbf{\hat{R}}$ that minimise reconstruction error on observed $\mathbf{Y}$. However, the exact problem of Boolean matrix factorisation is NP-complete, and its optimisation variant (i.e., minimising reconstruction error for a given rank) is NP-hard, necessitating heuristic or approximate methods in practice \cite{miettinen_recent_2020}.

\subsection{Related Work}

Given the combinatorial complexity of the problem, numerous heuristic and exact methods have been developed, each with different assumptions and optimisation strategies. In this section, we review key contributions in the literature on Boolean matrix factorisation and related techniques.



ASSO \cite{miettinen_discrete_2008} is a fast, greedy algorithm for Boolean Matrix Factorisation that constructs low-rank binary approximations by mining frequent itemsets from the input matrix. It identifies combinations of co-occurring features (called tiles) and selects a subset that best reconstructs the original matrix using Boolean OR and AND operations. ASSO iteratively selects itemsets that maximise coverage while minimising overlap and error, and solves a set cover problem to finalise the factor matrices, enabling interpretable and scalable BMF. MDL4BMF \cite{miettinen_mdl4bmf_2014} builds upon ASSO by integrating the Minimum Description Length (MDL) principle, which frames Boolean matrix factorisation as a model selection problem balancing complexity and goodness of fit. MDL4BMF includes a cost function that penalises the number of bits needed to encode the factor matrices $\mathbf{L}$ and $\mathbf{R}$. As a result, since the MDL objective balances model size and error, MDL4BMF can automatically select an appropriate number of factors $K$, avoiding the need to pre-specify $K$. PANDA+ \cite{lucchese_unifying_2014} is similar to MDL4BMF in that it generates candidate columns using a greedy algorithm; however, it directly optimises a complexity-based objective during the factorisation process, rather than applying complexity considerations post-hoc.


Using a continuous relaxation approach, PRIMP \cite{hess_primping_2017} formulates Boolean matrix factorisation as an optimisation problem with a two-part objective function. The first component is a reconstruction error term measured using standard algebraic matrix multiplication and the Frobenius norm, rather than Boolean algebra. Specifically, given an observed binary matrix $\mathbf{Y} \in \{0,1\}^{M \times N}$, PRIMP seeks factor matrices $\mathbf{L} \in [0,1]^{M \times K}$ and $\mathbf{R} \in [0,1]^{K \times N}$ by minimising
$\min_{\mathbf{L}, \mathbf{R}} \; \|\mathbf{Y} - \mathbf{L}\mathbf{R}\|_F^2 + \lambda \, \mathcal{R}(\mathbf{L}, \mathbf{R}),
$
where $\|\cdot\|_F$ denotes the Frobenius norm, $\lambda > 0$ is a regularisation parameter, and $\mathcal{R}(\mathbf{L}, \mathbf{R})$ is a regularisation term designed to promote binary-valued entries and low-rank structure in the factor matrices.

To solve this nonconvex and nonsmooth optimisation problem, PRIMP employs \textit{Proximal Alternating Linearized Minimization (PALM)} \cite{bolte_proximal_2014}, an iterative algorithm that generalises the Gauss-Seidel method to handle composite objective functions. PALM alternately updates $\mathbf{L}$ and $\mathbf{R}$ by performing proximal gradient steps that linearise the smooth part of the objective while incorporating proximal operators to handle nonsmooth regularisation terms. The proximal updates ensure convergence to a critical point despite the nonconvexity and nonsmoothness of the problem. By relaxing the binary constraints to continuous domains, PRIMP enables the use of efficient gradient-based optimisation methods, while the regularisation term encourages the learned matrices to be close to binary, facilitating interpretable factorisation. PRIMP evaluates the \textit{Minimum Description Length} (MDL) cost across multiple candidate values of $K$ for model selection.


\citet{kovacs_binary_2021} take a Mixed Integer Programming (MIP) approach for the BMF problem, leveraging the insight that a rank-$K$ matrix factorisation can be decomposed as the sum of $K$ rank-1 matrix factorisations. They construct a restricted master problem that iteratively selects the $K$ best rank-1 matrices from a potentially large set of candidate matrices. To efficiently handle this, they employ delayed column generation, dynamically adding advantageous rank-1 matrices to the master problem during optimisation. This technique improves scalability by avoiding the need to enumerate all candidates upfront; however, the approach is still limited to medium-sized datasets.
Another limitation of this approach is that the desired rank, $K$, must be prespecified before solving, requiring prior knowledge or additional model selection steps.

Of the limited more recent BMF methods that also estimate rank, few compare to the state of the art in literature, or do not significantly outperform them, please see Appendix \ref{sec:lit_rev} for a detailed comparison.

\subsection{Contributions}


We introduce a novel BMF approach, called \bfact, that solves a MIP related to BMF, selecting disjoint column sets that best explain the data. This can be combined with a second MIP-based or faster, yet effective, heuristic approach to perform standard BMF. The approach automatically selects the relevant rank using complexity measures or reconstruction error, and scales to large datasets. It performs well against existing BMF methods in simulated scenarios, standard real data benchmarks, and 14 scRNAseq datasets from the human cell lung atlas.

We provide the MIP formulation for finding an approximate BMF with disjoint column sets and provide the pricing problem to find the optimal disjoint BMF solution on smaller matrices, using delayed column generation. We provide a second MIP formulation that could be used for exact BMF on smaller datasets using delayed column generation. We show that complexity costs often used in BMF algorithms can favour sparser, higher-rank factorisations.

\section{Method}

\begin{figure}[!t]
    \centering
    \includegraphics[width=\linewidth]{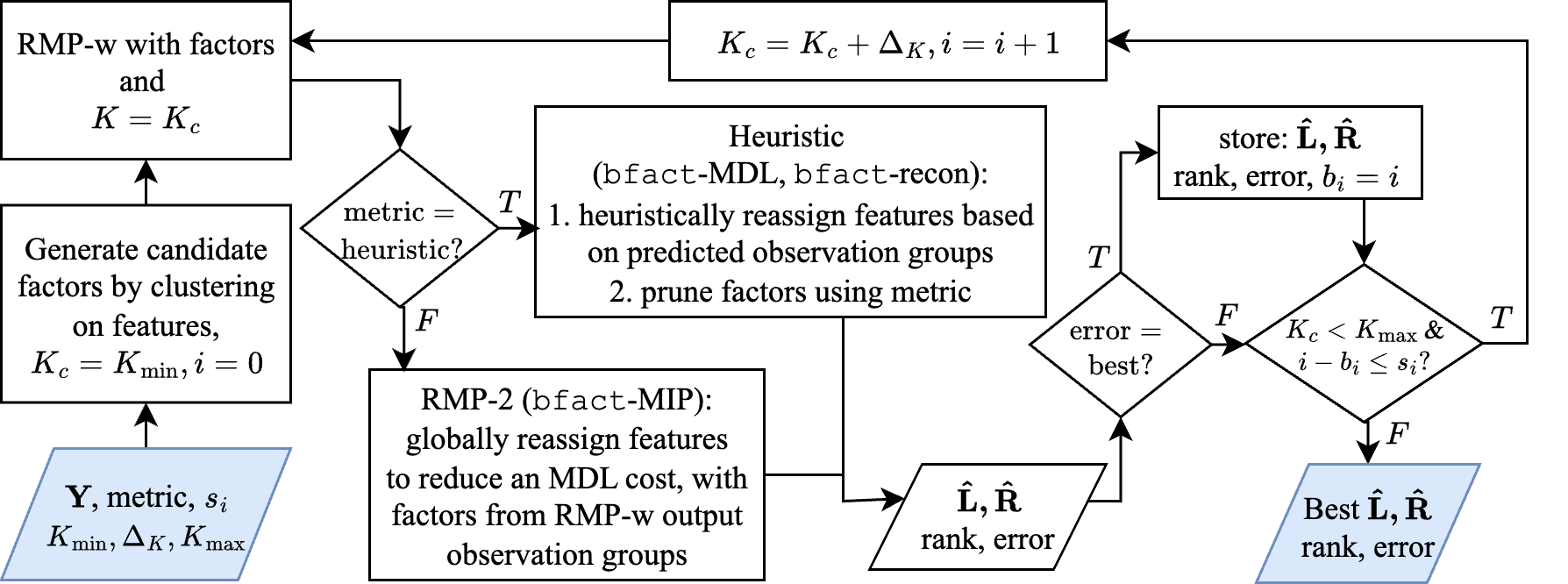}
    \caption{Overview of \bfact}
    \label{fig:schem}
\end{figure}

Our approach, illustrated in Figure \ref{fig:schem}, begins by generating candidate factors through clustering on features. We then solve a warm-started restricted master problem (RMP-w) that approximates the BMF by selecting up to $K_c$ of these factors ($K_c$ initialised to some $K_{\textrm{min}}$). Depending on the selected metric, the method either heuristically reassigns features and prunes factors (\bfact-recon or \bfact-MDL) or performs a second combinatorial approach to refine the factorisation (\bfact-MIP). The process iteratively increases the maximum number of factors, $K_c$, stopping to give the best factorisation solution if the metric error does not improve within $s_i$ steps. This two-stage framework—starting from disjoint candidate factors and refining via heuristic or optimisation—echoes the ASSO methodology while incorporating modern optimisation techniques.
Below, we explain each of these steps in detail.

\subsection{Master problem for approximate BMF}

Here we borrow ideas from the combinatorial optimisation world.
We define a `master problem' (MP) that approximates the BMF by finding sets of (mostly) disjoint features that best explain the observations. 
For our observed binary matrix $\mathbf{Y}$, consider the set $\cal A$ of all possible non-zero feature-sets, or factors, $\alpha$, where $\delta_\alpha \in \{0, 1\}^N$ and $|{\cal A}| = 2^N - 1$. 
We define also $c_{i, \alpha}$, the cost for observation $i$ of choosing factor $\alpha$, as:
\begin{align}    
C_{i, \alpha} &= \left ( \sum_{j| (i,j) \in E} \delta_{\alpha, j}, \sum_{j| (i,j) \notin E}  \delta_{\alpha,j} \right ) \label{eq:cost} \\
c_{i, \alpha} &= \min C_{i, \alpha} \\
l_{i, \alpha} &= \arg \min C_{i, \alpha}
\end{align}
where $E = \{(i, j) | Y_{i,j} = 1\}$. The left term of $C_{i, \alpha}$ is the intersection of features in an observation and in a factor. While the right term can be thought of as the complement of an observation - that is, the features included in a factor that are not present in the observation.   

The initial MP is thus defined as:
\begin{align}
    \min \sum_{j} & u_j \sum_i Y_{ij} + \sum_{\alpha}z_\alpha\sum_{i}c_{i, \alpha} \label{eq:obj} \\
    \sum_{\alpha \in {\cal A}} & z_{\alpha} \leq K 
    \label{eq:con_pp2} \\
    \sum_{\alpha \in A} & \delta_{\alpha, j}z_{\alpha} + u_j \geq 1 \label{eq:con_pp1} 
\end{align}
where $u_j \in \mathbb{R}^+, \forall j = 1, \dots, N$, and $z_\alpha \in \{0, 1\}\; \forall \alpha \in {\cal A}$.

To interpret this formulation, first consider if Equation \ref{eq:con_pp1} were an equality (i.e $\sum_{\alpha \in A} \delta_{\alpha, j}z_{\alpha} + u_j = 1$).
This enforces that no factors may share overlapping features, making them disjoint, and here $u_j$ represents features that are not included in any factor.
In this case, Objective \ref{eq:obj} is the reconstruction error, where the left term penalises false-negatives for any features that are omitted from the chosen factor representation.
All other features are included in a factor, and hence, for a given chosen factor, each sample either:
\begin{enumerate}
    \item Uses that factor - if the factor includes features not in the sample, then costs associated with false-positives should be incurred (right term of Equation \ref{eq:cost}).
    \item Does not use that factor - any features included in that factor cannot be used in another factor (assuming equality of Equation \ref{eq:con_pp1}), and overlapping features in the sample will not be covered, hence costs associated with false-negatives should be incurred (left term of Equation \ref{eq:cost}).
\end{enumerate}

Although the inequality in Equation \ref{eq:con_pp1} relaxes the enforcement of disjoint factors, the precomputed cost still implicitly assumes disjointness in its calculation of false positives and negatives, meaning false negatives induced from one chosen factor across samples will still be penalised even if another factor covers those features.
Similarly, false positives will be penalised according to standard (not Boolean) addition.
Hence, the formulation is still mostly disjoint; the constraint relaxation makes it easier to solve (particularly in the RMP-w setting introduced in the next setting).
Although a non-disjoint formulation can be proposed (and is proposed later in RMP-2), this formulation excels because costs are precomputed per factor instead of relying on costs dependent on a composition of chosen factors that cannot be precomputed at scale.
Finally, Equation \ref{eq:con_pp2} allows at most $K$ factors to be selected.

Hence, the solution to the MP above gives an approximate, (mostly) disjoint, BMF with $\mathbf{\hat{R}}_d = \{\delta_{\alpha, j} | z_\alpha = 1\}$, and $\mathbf{\hat{L}}_d = \{l_{i, \alpha} | z_\alpha = 1 \}$.
Note, the disjoint factorisation is similar to a clustering approach on the features, however unlike in clustering, the term $u_j$ allows any number of features to be excluded from selected factors.
See the Appendix \ref{sec:ill_eg}, Figure \ref{fig:ill_eg}, for a brief example illustrating the formulation.

\subsection{Warmstarted restricted master problem (RMP-w)}
Solving the MP above would require enumerating all possible factors ${\cal A}$, exponential in the number of variables and intractable.
Delayed column generation is an optimisation technique that can combat this and involves solving the restricted master problem (RMP) \cite{dantzig1960decomposition}. 
The RMP is the master problem `restricted' to a subset of possible factors, ${\cal A'} \subseteq {\cal A}$, where typically $|{\cal A'}| \ll |{\cal A}|$, making it tractable to solve.

In delayed column generation, factors that will reduce the objective are iteratively added to the RMP by solving the pricing problem, based on dual variables from the RMP linear relaxation \cite{vanderbeck_generic_2006}. Eventually, no more factors (i.e. columns) can be added that will reduce the objective, and the global optima of the linear relaxation is found, without having to realise the full MP linear relaxation. To solve the integer MP to optimality, a branch-and-price approach can be taken, again without having to realise the full MP.

Such an approach can work even when the RMP starts with an empty factor set.
However, it is often faster to warmstart the RMP with \emph{a priori} candidate factors.
A benefit of the disjoint formulation is that clustering on the features provides a simple way to generate good candidate factors. 
Hence, we perform hierarchical and Leiden clustering across the features to generate a set of candidate factors, which, together with our RMP, we call RMP-w.
For exact details on the candidate factor generation, see Appendix \ref{sec:cand_gen}

We first attempted to solve the RMP-w using a delayed column generation approach, like \citet{kovacs_binary_2021}. 
Here, we found that the pricing problem struggled to reduce the linear relaxation objective in a reasonable time frame (see Appendix \ref{sec:col_gen}), suggesting that the solution to RMP-w already produces good-quality disjoint factorisations.
Given the formulation is an approximation to the BMF, we proceed using RMP-w alone (i.e. using only the cluster candidate factors, without solving to optimality).

Note, our RMP scales (variables, constraints) with $(|{\cal A'}| + N,\;N)$. This makes it more computationally tractable than an exact approach such as \citet{kovacs_binary_2021}, which scales with $(\rho MN + |{\cal D'}|,\; \rho MN)$, where ${\cal D'}$ is the restricted set of all rank-1 $M \times N$ matrices, and $\rho$ is the density of ones in the data matrix. 

\subsection{Two-step BMF}

RMP-w does not directly solve for a BMF, however, we hypothesised that it may provide a good basis for a BMF after adding some post-processing. Hence, we adopt a two-step approach. Intuitively, the first step, RMP-w, selects (mostly) disjoint pregenerated feature sets to find likely observation sets, and the second step uses these observation sets to update the feature sets, allowing them to share features. The second step also controls the model complexity to select lower-rank approximations, where appropriate. 

\subsubsection{Two-step MIP-BMF:} 

Here, we formulate a second restricted master problem (RMP-2), similar to the above, but instead for (nearly) exact BMF.
Consider the set of all factors $\cal B$, now with the set of observations,  $\delta_\beta \in \{0, 1\}^M$ associated with each factor, $\beta \in {\cal B}$.
We define RMP-2 to be over a restricted factor set ${\cal B'} \subseteq {\cal B}$, given by:
\begin{align}
\min \sum_{(i,j) \in E} (1 - e_{i,j})   &+ \sum_{(i,j) \notin E} \sum_{\beta \in {\cal B'}} \delta_{\beta, i} R_{\beta, j} \notag \\
+\sum_{\beta \in {\cal B'}} \sum_{i=1}^{M} z_\beta \delta_{\beta, i} &+ \sum_{\beta \in {\cal B'}} \sum_{j=1}^{N}R_{\beta, j}   \label{obj:rmp2}
\end{align}
subject to
\begin{align}
\sum_\beta & \delta_{\beta, i}  R_{\beta, j} \geq e_{i, j},  \quad \forall \; i, j \in E \label{eq:rmp2_con1} \\
\sum_j & R_{\beta, j} \leq N z_\beta, \quad \forall \; \beta \in {\cal B'}  \\
\sum_\beta & z_\beta \leq K
\end{align}
where $z_{\beta} \in \{0, 1\}, \forall \; \beta \in {\cal B'}$, $
R_{\beta, j}  \in \{0, 1\}, \forall \; \beta \in {\cal B'},\; j = 1 \dots N$ and $e_{i, j}  \in \{0, 1\}, \forall \; i, j \in E$.
Here again $E = \{(i,j) | Y_{ij} = 1\}$.

The first two terms in objective \ref{obj:rmp2} capture false negatives and positives, respectively. The third and fourth terms are proxies for the complexity of the model and correspond to $|\mathbf{\hat{L}}|$ and $|\mathbf{\hat{R}}|$, these regularise the model to select fewer factors, and sparser $\mathbf{\hat{L}}$ and $\mathbf{\hat{R}}$ (derived from chosen factors). 
This is similar to the regularisation taken in PANDA+ \cite{lucchese_unifying_2014}.
The third and fourth terms could be dropped to give a BMF for a specific $K$.

Equation \ref{eq:rmp2_con1} uses the term $e_{i,j}$ to allow for boolean logic, restricted to non-zero elements of $Y$ to reduce the number of variables in the model.
Hence, Boolean logic is encoded only for true positives, so overlaps at false positives are penalised more than in exact BMF.
This could easily be remedied by using $e_{i,j}$ for all elements in the matrix, at the expense of scalability.

RMP-2 scales (variables, constraints) with $(N(\rho M + |{\cal B'}|), \;\rho MN + |{\cal B'}|)$ where $\rho$ is the density of the matrix.
A column generation approach could be used for RMP-2, but it is likely to be intractable for larger problems, given its scaling and that of its associated pricing problem.
It could also be solved similarly to RMP-w, using candidate factors. 
However, using candidate factors generated by a clustering approach would not capitalise on the added expressiveness of this formulation, given that such candidates are disjointly derived (though their composition may not be).

Instead, we opt to use RMP-2 as a second step after finding a solution to RMP-w, to refine the feature sets of factors (i.e. $\mathbf{\hat{R}}_d$) and chosen factors.
Again we use candidate factors by taking ${\cal B'}$ as the unique set of observation memberships across the selected factors of the solved RMP-w, with $\delta_\beta, \; \beta \in {\cal B'}$ given by the unique columns of $\mathbf{\hat{L}}_d$. 
Hence, RMP-2 selects known observation factor groups found with RMP-w, and reassigns features to them based on these groups, removing any redundant factors through the regularisation terms.

Given $|{\cal B'}| \leq K$ in our RMP-2 approach, we found it was still computationally tractable for larger matrices (order 100k x 20k, $\rho \approx 0.1$), but with heavy memory requirements and longer runtimes.
Adapting the above formulation to be exact (given the factors) would require $e_{i,j}$ to cover all locations, intractable given an approximate 10-fold increase in the number of variables.
Given these, we explored a less memory-intensive, heuristic approach. 

\subsubsection{Two-step Heuristic-BMF}

By grouping observations containing the same factor(s), we can recover additional features present in the grouping that should be added to that factor's feature set.
RMP-2 does this globally based on the defined observation sets and using a basic complexity cost.

To allow features to be allocated to multiple factors, a coarse grid search is performed to identify at which global proportion of representation in an observation group a feature should be added to a factor.
This optionally uses either the reconstruction error or MDL loss as defined in \citet{hess_primping_2017}, see Appendix \ref{sec:mdl}, which accounts for the sparsity, hence implicitly the rank, of the factorisation.
The approach is formalised in Algorithm \ref{ref:algo_1}.

\begin{algorithm} [!t]
\caption{Reassign Features}\label{ref:algo_1}
\begin{algorithmic}[1]
\Function{reassign}{$Y, L,\hat{R}, \text{metric}, \Delta t$}
    \State $I_{kj} =\sum_{i} L_{ik}Y_{ij}$; \quad $I \gets \{ I_{kj} \}_{K\times N}$\;
    \State $N_k$ = $\sum_{i} L_{ik}$; \quad $N \gets \{ N_k \}_K$\;
    \State $R_b \gets \hat{R}$
    \State $E \gets \textsc{error}(Y, L, \hat{R}, \text{metric})$
    \For{$t = 0$ \textbf{to} $1$ \textbf{step} $\Delta t$}
    \State $R_t = (I > \frac{1}{2}N(1 + t))\ \lor\  \hat{R} $
    \State $E_t = \textsc{error}(Y, L, R_t,\text{metric})$
    \If{$E_t < E$} 
    \State $R_b \gets R_t; \quad E \gets E_t$
    \EndIf
    \EndFor
  \State \Return $R_b$
  \EndFunction
\end{algorithmic}
\end{algorithm}

Following the reassignment of features, we greedily remove redundant factors, or factors that result in marginal improvement of the loss.
To remove a factor with reconstruction error, the error without the factor must be some minimum percentage $f$ of the reconstruction error with the factor.
The MDL loss already accounts for removing a factor through reduced complexity (so $f=1$ with MDL loss).
This is formalised in Algorithm \ref{ref:algo_2}.

\begin{algorithm}[t] 
\caption{Iteratively remove factors}\label{ref:algo_2}
\begin{algorithmic}[1]
\Function{prune}{$Y, L, R, \text{metric}, f=1$}
    \State $M, K \gets \text{dim}(L)$
    \State $E \gets \textsc{error}(Y, L, R)$
    \While{$K > 0$}
        \State $E_\text{ls} = []$
        \For{$r = 1 \dots K$}
            \State $L_c \gets L, R_c \gets R$
            \State $L_c[:, r] \gets 0, R_c[r,:] \gets 0$
            \State $E_\text{ls}.\text{append}(\textsc{error}(Y, L_c, R_c, \text{metric}))$
        \EndFor
        \State $b_c \gets \arg \min E_\text{ls}$
        \If{\textbf{not} $ E_\text{ls}[b_c] \leq f E$}
        \State \textbf{break} // no better sln
        \EndIf
        \State $L \gets L.\text{delete}(\text{col}=b_c)$
        \State $R \gets R.\text{delete}(\text{row}=b_c)$
        \State $K \gets K - 1$
        \State $E \gets E_{ls}[b_c]$
    \EndWhile
  \State \Return $L, R$
 \EndFunction
\end{algorithmic}
\end{algorithm}

\subsection{Finding optimal rank}

Under the disjoint factorisation master problem, the model may achieve a lower overall objective when $K$ is overspecified, see Figure \ref{fig:ill_eg}. Although the postprocessing should combine or remove redundant features, the initial rank specification will affect downstream reconstruction, reassignment and feature pruning. Hence, we perform the process over multiple initial ranks and select the best result (noting that the RMP-w can be initialised once, and updated with different $K$ values, after which it is fast to solve). If increasing the rank does not result in a better solution for $s_i$ iterations, the algorithm is stopped early. 
This is formalised in Algorithm \ref{ref:algo_3}.

We term \bfact as the sequential pipeline of w-RMP followed by the second step (RMP-2 or heuristic) followed by algorithm 3.
\bfact-MIP is this pipeline where the second step is RMP-2 (as w-RMP and RMP-2 are both MIPs).
\bfact-recon is the pipeline where the second step is heuristic (algorithms 1 and 2) with reconstruction error, while \bfact-MDL is the same using the MDL cost.

\begin{algorithm} 
\caption{Select best rank over multiple K}\label{ref:algo_3}
\begin{algorithmic}[1]
\Function{pipeline}{$Y, K_{\min}, K_{\max},\Delta K, \text{metric}, f=1, s_i=2$}
    \State $A = \textsc{gencols}(Y)$
    \State $V_{b} = \text{null}, \ E_b = \text{null}, \ K_b = \text{null}$
    \State $b_i=0, i=0$
    \For{$K_c = K_{\min}$ \textbf{to} $K_{\max}$ \textbf{step} $\Delta K$} 
        \State $L, R = \textsc{RMP}_1(Y, A, K_c)$
         \If{$\text{metric}$ \textbf{is} $\text{mip}$} 
            \State $L, R \gets \textsc{RMP}_2(Y, L, K_c)$ 
        \Else
            \State $R \gets \textsc{reassign}(Y, L, R, \text{metric})$ 
            \State $L, R \gets \textsc{prune}(Y, L, R, \text{metric}, f)$ 
        \EndIf
        \State $E_k \gets \textsc{error}(Y, L, R, \text{metric})$
        \If{$E_b$ \textbf{is} $\text{null}$ \textbf{or} $E_k \leq f^{K_c - K_b}E_b$ } 
            \State $b_i \gets i,E_b \gets E_k,\ V_b \gets (L, R),\ K_b \gets K_c$
        \EndIf
        \If{$i - b_i > s_i$} 
            \State \textbf{break} // stop early
        \EndIf
        \State $i += 1$
        \EndFor
  \State \Return $V_b$
 \EndFunction
\end{algorithmic}
\end{algorithm}

\section{Experiments}

\begin{figure*}[h!]
    \centering
    \includegraphics[width=\linewidth]{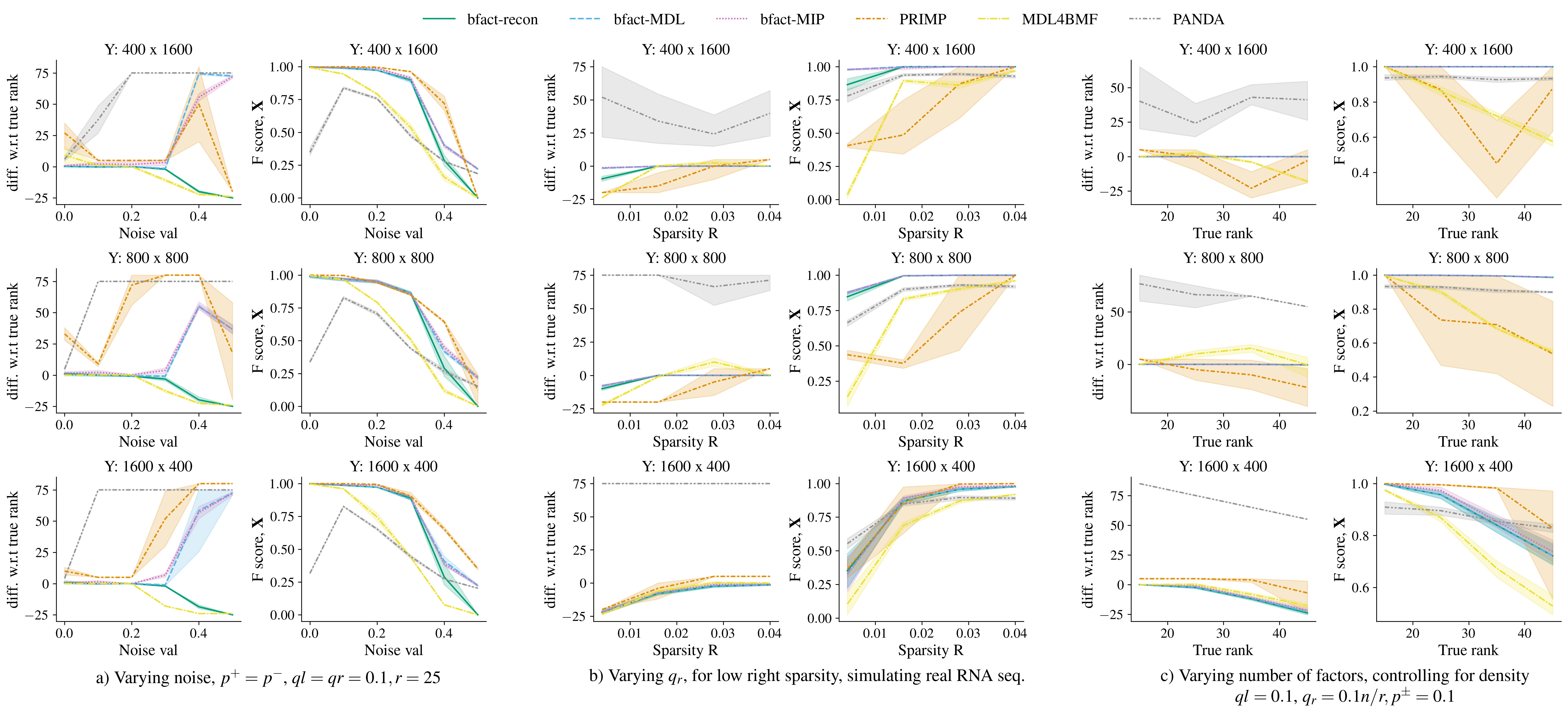}
    \caption{Simulation results.} \label{fig:sim_results}
\end{figure*}

\begin{figure*}[h!]
    \centering
    \begin{subfigure}{0.65\linewidth}
         \includegraphics[width=\linewidth]{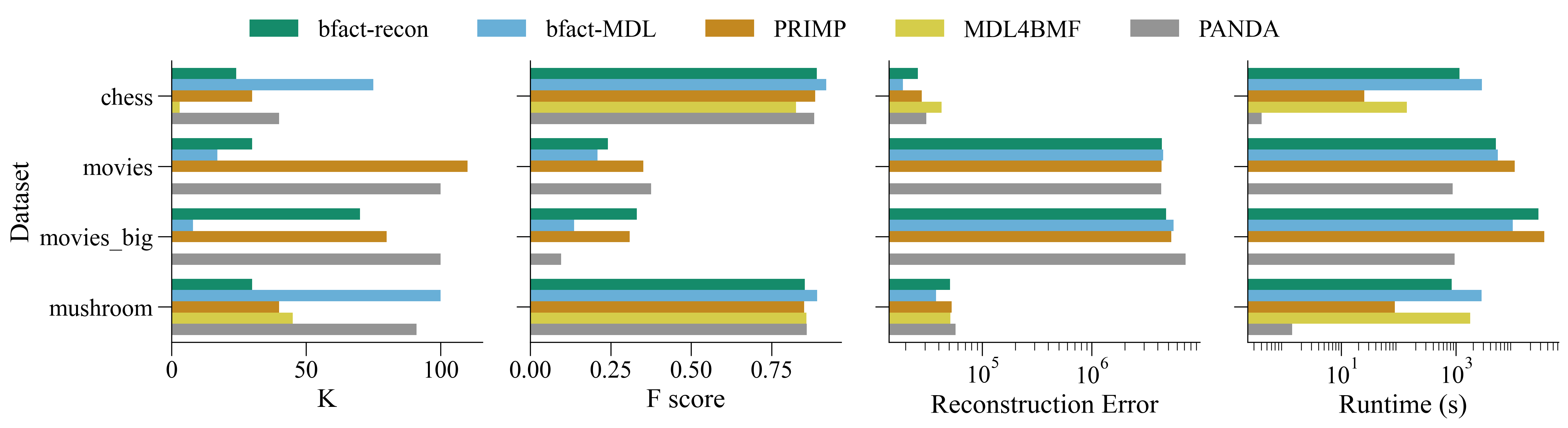}
    \end{subfigure}
    \begin{subfigure}{0.33\linewidth}
        \centering
        \includegraphics[width=0.8\linewidth]{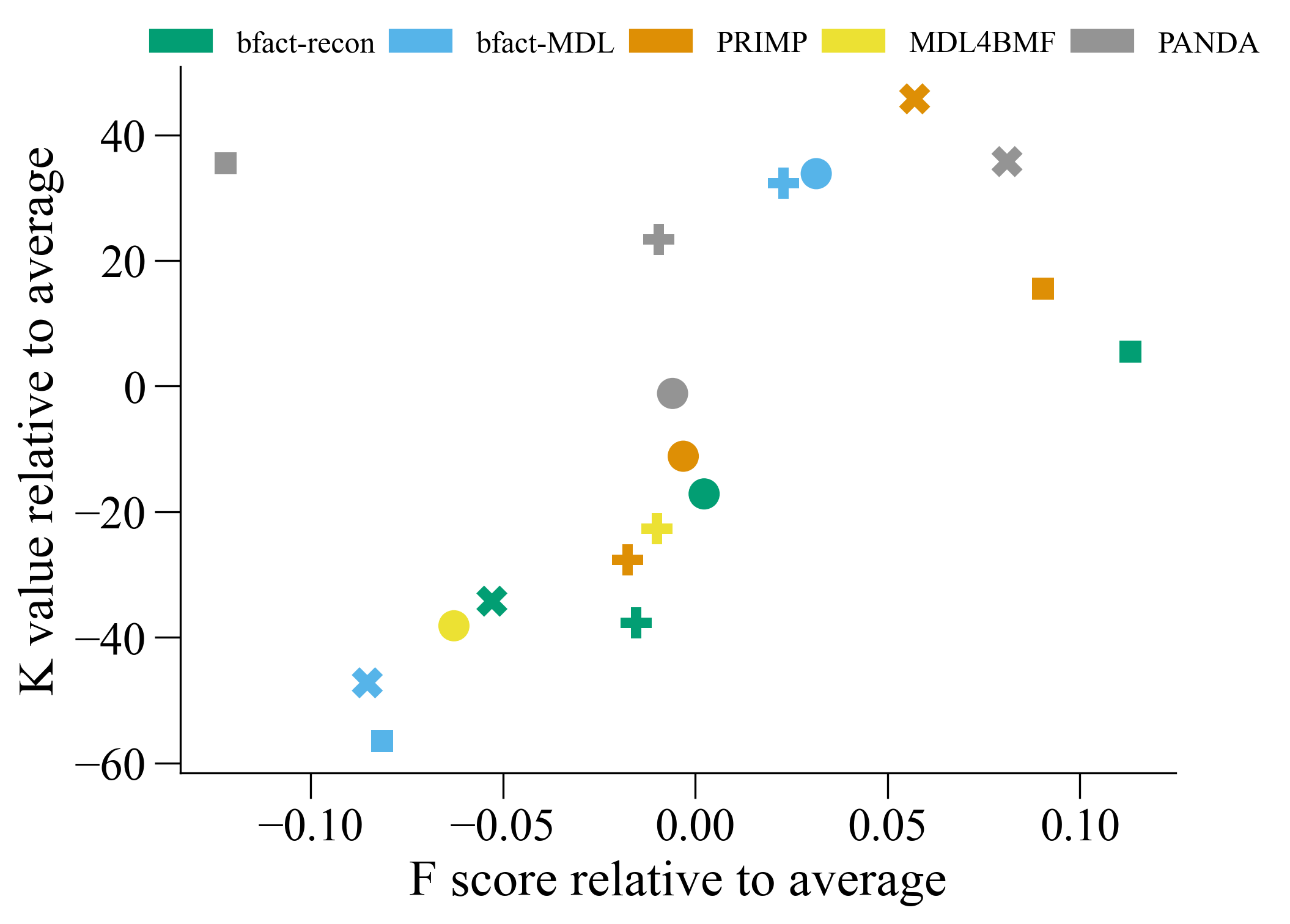}
    \end{subfigure}
    \centering
    \begin{subfigure}{0.65\linewidth}
         \includegraphics[width=\linewidth]{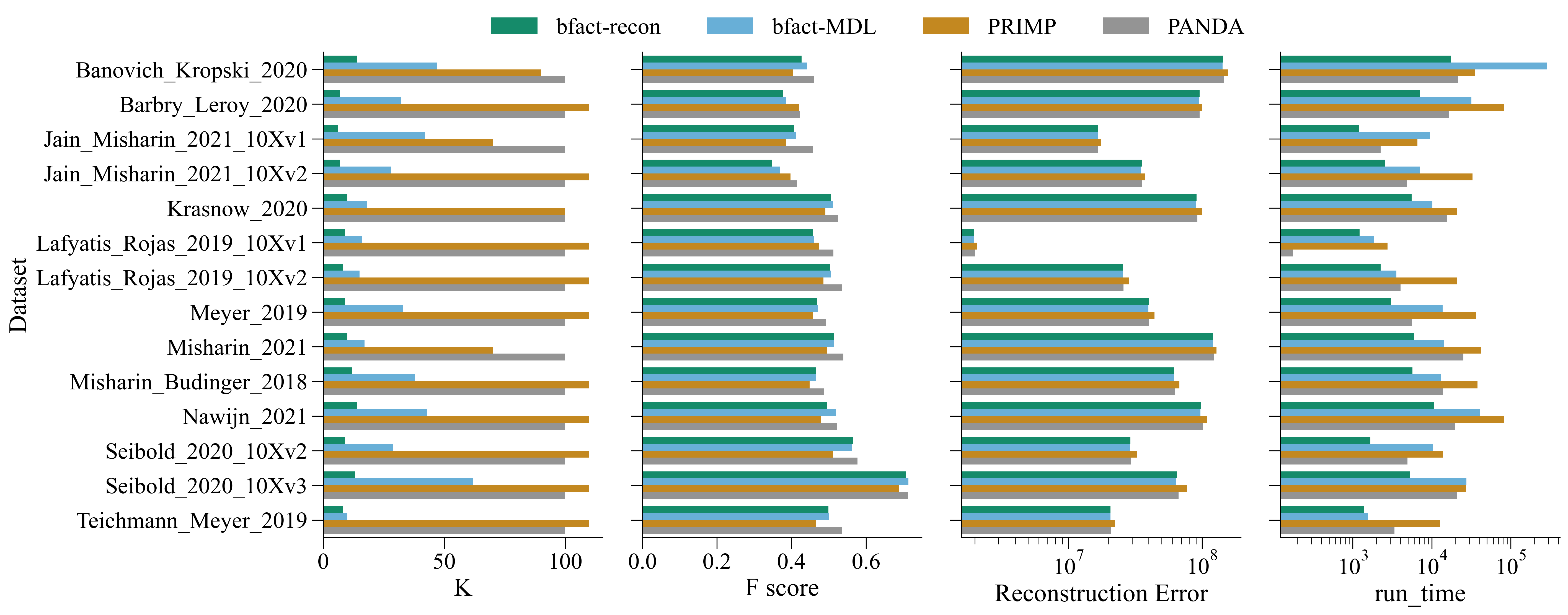}
    \end{subfigure}
    \begin{subfigure}{0.33\linewidth}
        \includegraphics[width=\linewidth]{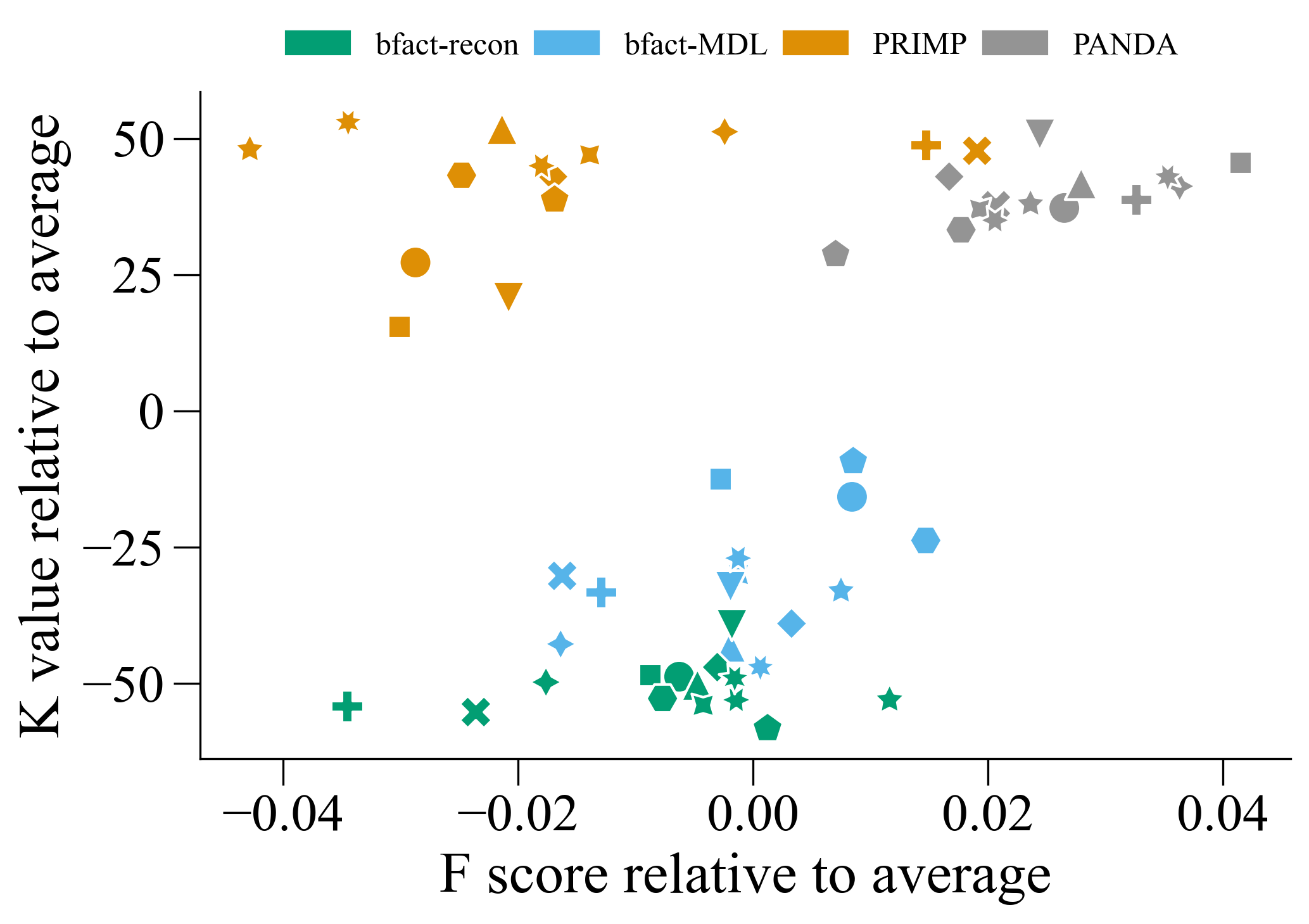}
    \end{subfigure}
    \caption{Real benchmarking results, standard benchmarks (top), and HLCA RNA-sequencing results (bottom). Shapes on right hand plots correspond to different datasets.} \label{fig:real_results}
\end{figure*}

Here, we compare \bfact to existing approaches PRIMP, PANDA+ and MDL4BMF. 
For implementation details, please see Appendix \ref{sec:meth_imp}.

\subsection{Data}

\paragraph{Simulation.}

We considered several simulation setups to benchmark our method. We generated a variety of scenarios with different matrix sizes, underlying ranks, noise levels and data density scenarios. Some rows and columns were also generated to be `nuisance' variables, imitating realistic datasets where some features or observations are not relevant to the factorisation. 

Formally, the main simulation set-up is as follows:
\begin{align*}
    \text{inputs:} \quad M, N, k, q_l, &q_r, v_i, v_j, p^+,p^- \\
    \mathbf{L} \sim \{\text{Ber}(q_l)\}_{M\times k} &\qquad 
    \mathbf{R} \sim \{\text{Ber}(q_r)\}_{k\times N} \\
    n \sim \{\text{Ber}(v_i)\}_{M} &\qquad
    m \sim \{\text{Ber}(v_j)\}_{N} \\
    \mathbf{L}[n > 0, .] = 0 &\qquad
    \mathbf{R}[.,m > 0] = 0 \\
     \mathbf{X} = \mathbf{L}\mathbf{R} &\qquad
     \bfeps^{\pm} \sim \{\text{Ber}(p^\pm)\}_{M\times N}\\
    \mathbf{Y} = \mathbf{X} + {\bfeps}^+[&\mathbf{X} = 0] - {\bfeps}^-[\mathbf{X} > 0]
\end{align*}
where Ber$(\alpha)$ represents the Bernoulli distribution with probability $\alpha$.
For each simulated experiment, we take 5 replicates.
\medbreak
Note, it is possible that for extremely sparse sampling the true rank is lower than specified; however, we assume the effect of this is negligible and would likely be captured by measured algorithms.

\paragraph{Real-world datasets.}

We also consider real-world data sets. This includes binarised versions of the Chess and Mushroom UCI datasets \cite{markelle_kelly_uci_nodate} and two versions of the MovieLens 10M dataset, \cite{harper_movielens_2015}, where rows are users and columns are movies, with entries the star rating given by a user to a movie. Following \citet{hess_primping_2017}, we set $Y_{ij} = 1$, if a user rates a movie with more than 3 stars. This constitutes the larger dataset, we then also take a smaller dataset filtered to select users who recommend more than 50 movies and movies that receive at least 5 recommendations.\footnote{While we follow precedent in previous publications, in some of the real data examples, the data has been binarised from discrete, categorical data using one-hot encoding. While such transformations enable the use of binary matrix factorisation, it is unclear whether they truly reflect the nature of the data.} 

We also use single-cell RNA sequencing data from the Human Lung Cell Atlas (HLCA), \cite{sikkema_integrated_2023}, which consists of 14 separate datasets on lung-derived cell types. We used the clean raw counts for each dataset and binarised them based on zero/non-zero values in the data. We also removed genes that were not expressed in at least 0.5\% of cells, and cells that expressed fewer than 200 genes and more than 10,000. The size and density of each example are included in Table \ref{tab:dataset_stats}.

\begin{table}[h]
\centering
\setlength{\tabcolsep}{1mm} 
{\fontsize{9}{11}\selectfont
\begin{tabular}{>{\raggedright\arraybackslash}p{.9cm}cccc}

\hline
\textbf{Origin} & \textbf{Dataset} & \textbf{M} & \textbf{N} & \textbf{Density} \\
\hline
UCI & Chess & 3196 & 75 & 0.493 \\
UCI & Mushroom & 8124 & 119 & 0.193 \\
Movie Lens & Movies & 29 980 & 9144  & 0.018 \\
Movie Lens & Movies Big & 69 878 & 10 677  & 0.008 \\
\hline
HLCA & Banovich Kropski 2020 & 121894 & 14495 & 0.101 \\
HLCA & Barbry Leroy 2020 & 74484 & 15047 & 0.102 \\
HLCA & Jain Misharin 2021 10Xv1 & 12422 & 13423 & 0.124 \\
HLCA & Jain Misharin 2021 10Xv2 & 33135 & 13392 & 0.094 \\
HLCA & Krasnow 2020 & 60982 & 15139 & 0.133 \\
HLCA & Lafyatis Rojas 2019 10Xv1 & 2921 & 11943 & 0.073 \\
HLCA & Lafyatis Rojas 2019 10Xv2 & 21258 & 13818 & 0.117 \\
HLCA & Meyer 2019 & 35554 & 14153 & 0.103 \\
HLCA & Misharin 2021 & 64842 & 15938 & 0.157 \\
HLCA & Misharin Budinger 2018 & 41219 & 14057 & 0.136 \\
HLCA & Nawijn 2021 & 70395 & 15579 & 0.119 \\
HLCA & Seibold 2020 10Xv2 & 12127 & 15718 & 0.215 \\
HLCA & Seibold 2020 10Xv3 & 21466 & 17825 & 0.310 \\
HLCA & Teichmann Meyer 2019 & 12231 & 14855 & 0.150 \\
\hline
\end{tabular}
}
\caption{Dataset statistics}
\label{tab:dataset_stats}
\end{table}

\subsection{Evaluation metrics}
To evaluate each approach we compared the predicted rank to the true underlying rank, and the F1 score of the predicted signal matrix $\mathbf{\hat{L}}\mathbf{\hat{R}}$ with respect to the true signal matrix $\mathbf{X}$ in simulated settings, or the observed data matrix, $\mathbf{Y}$ in real data. 
See Appendix \ref{sec:eval_met} for details of the F1 score.
For real data, we also present the reconstruction error, given by $\sum_{ij}  |\hat{X}_{ij} - Y_{ij}|$.

Methods with higher-rank predictions can overfit to noise, which could give a higher F1 score on $\mathbf{Y}$ but a low F1 score on the true signal matrix $\mathbf{X}$.
Similarly, some methods may split factors into smaller ones, obscuring the true decomposition to give a high F1 score (even on $\mathbf{X}$) but also a higher predicted rank than necessary.
Hence, the predicted rank should be considered in tandem with the reconstruction and $F1$ score.
Other approaches compare MDL terms, which aim to balance model complexity and reconstruction error implicitly. 
However, as we show in Appendix \ref{sec:mdl_lim}, MDL costs favour sparse decomposition, which do not necessarily align with a BMF, particularly for lower-rank, higher-complexity factorisations.


\subsection{Results}

\paragraph{Simulations.} Figure \ref{fig:sim_results} and Figures \ref{fig:app_sim_results_A}, \ref{fig:app_sim_results_B} (Appendix \ref{sec:ext_res}) show that the three variations of \bfact perform similarly in both F1 score and rank estimation across different simulated regimes. PANDA+ consistently overestimates rank and achieves lower F1 scores but it should be noted that it requires less computational time than other methods (Figure \ref{fig:sim_time}). PRIMP does particularly poorly when the sparsity of $\mathbf{R}$ is low. MDL4BMF does well at estimating rank but has lower F1 scores. It is much slower than other methods to run despite being provided double the number of CPUs (Figure \ref{fig:sim_time}). All methods perform worse at higher density.
On simulated data, \bfact has a comparable run-time to PRIMP (Figure \ref{fig:sim_time}). Interestingly, \bfact-MIP does slightly worse at recovering the true rank (Figures \ref{fig:app_sim_results_A}, \ref{fig:app_sim_results_B}). 
Likely, this is due to the regularisation approach taken, $|\mathbf{\hat{L}}| + |\mathbf{\hat{R}}|$, which encourages sparse, not necessarily low-rank reconstruction. This is supported by the fact that it has as high F-scores as the other \bfact approaches. 
Given \bfact-MIP is also slower and more memory-intensive, we do not explore \bfact-MIP further. For \bfact-recon and \bfact-MDL, the time-limiting factor is the heuristic post-processing, rather than the combinatorial RMP-w problem.

\paragraph{Real-world data.}
On real-world benchmark data, Figure \ref{fig:real_results} (top) shows that both \bfact variants achieve comparable F-scores and reconstruction errors to other methods, and in particular, \bfact-recon does this with fewer factors (K).
A potential reason for this is that \bfact-recon is the only method not to use a complexity-based score for rank approximation, and such scores do not always favour the lowest rank factorisation (see Appendix \ref{sec:mdl_lim}).
Furthermore, some of these datasets have been one-hot encoded from categorical data; hence, it is unclear whether a BMF is the correct model for such data.
On the 14 single-cell RNA sequencing datasets, for which there is more precedent for a BMF, Figure \ref{fig:real_results} (bottom) shows the \bfact variants achieve both performant F-scores and reconstruction errors but use significantly fewer factors than other methods.
In scRNA-seq, PANDA+ is likely overfitting to noise rather than signal, given its high predicted rank.
To support this, consider simulated results in Figure \ref{fig:app_sim_results_B}, where PANDA+ achieves the highest F score on the observed data matrix $\mathbf{Y}$, despite having the lowest F score on the signal matrix $\mathbf{X}$, also with a very high rank.
Despite some promising results on lower-dimensional standard datasets, MDL4BMF often took too long to run ($>$2 days with 24CPUs), so it has not been included for the corresponding datasets.

\section{Conclusion}

We present a new binary matrix factorisation approach, including several sub-variants, called \bfact, which uses a hybrid combinatorial optimisation approach based on \emph{a priori} factors generated from existing clustering algorithms. We show it performs well in simulations and particularly when applied to single-cell RNA sequencing data from the Human Lung Cell Atlas. We demonstrate that it is scalable to data of this size. We also include formulations and the pricing problem for finding the optimal disjoint BMF on smaller datasets.
\bfact is available as a pip installable package. 

\section{Acknowledgements}

CY is supported by an UKRI Turing AI Acceleration Fellowship (Ref: EP/V023233/1) and EPSRC grant (Ref: EP/Y018192/1). EV is supported by the Oxford EPSRC Centre for Doctoral Training in Health Data Science (Ref: EP/S02428X/1).

\bibliography{aaai2026}


\appendix
\raggedbottom
\setcounter{figure}{0}
\setcounter{equation}{10}
\setcounter{secnumdepth}{2} 
\setcounter{section}{1} 
\renewcommand{\thefigure}{A\arabic{figure}}
\section*{Appendix}

\begin{figure*}[h!]
    \centering
    \includegraphics[width=\linewidth]{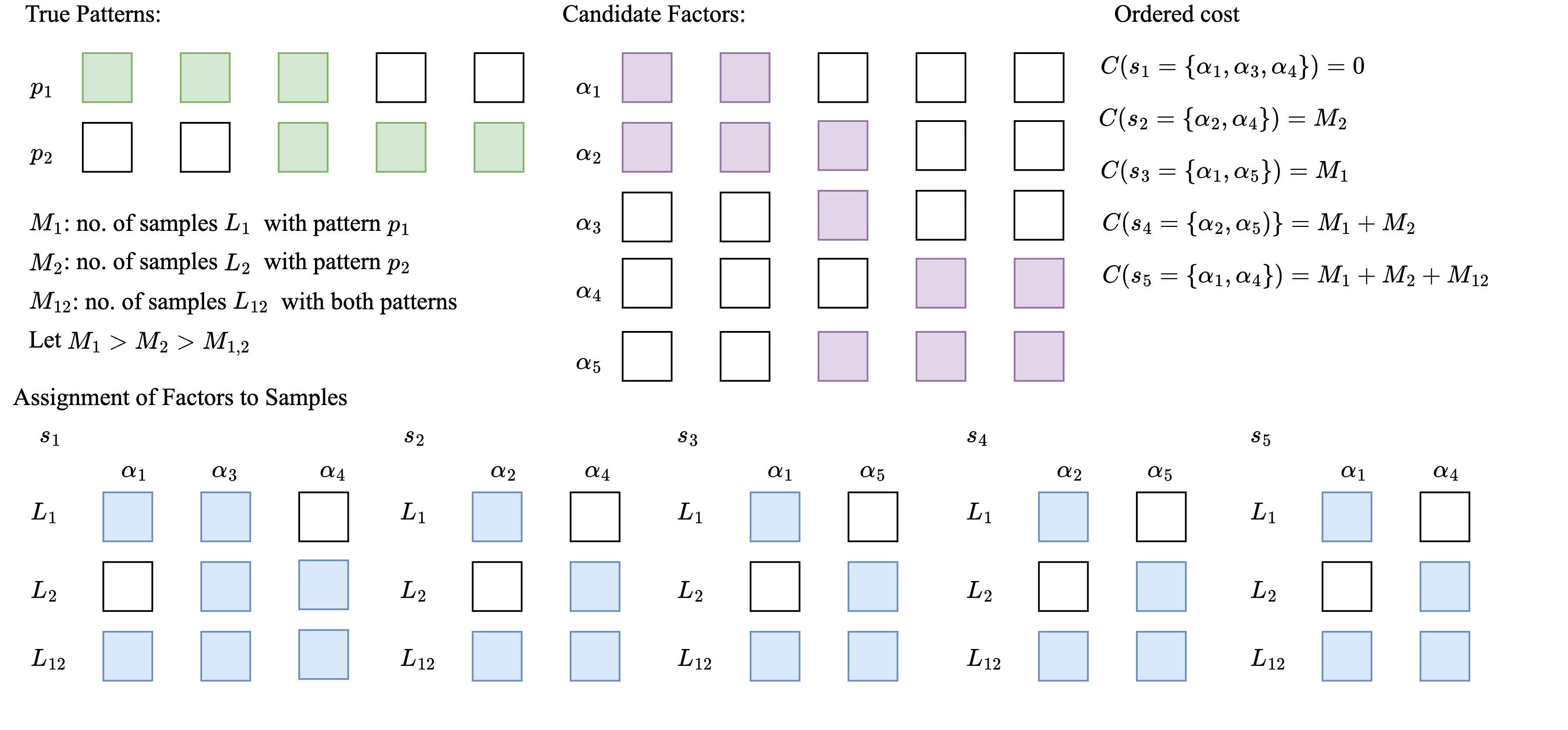}
    \caption{Illustrative example of RMP-1 cost.}
    \label{fig:ill_eg}
\end{figure*}

\subsection{Related work further details}\label{sec:lit_rev}
In the literature, many existing approaches have been proposed for Boolean matrix factorisation.
However, few incorporate optimal selection of rank, $K$, and often these methods are developed in the context of small to medium scale matrices and do not offer runtimes practical for scRNA-seq datasets.
As the problem is NP-hard, there is often a trade-off between exact combinatorial approaches limited to smaller datasets and approximate approaches that can scale to larger datasets.
Table \ref{tab:bmf_comparison} introduces existing approaches and details why they were or were not implemented in the work.
PRIMP has been shown in numerous works to perform well, hence newer works would ideally compare to PRIMP, where relevant \cite{dalleiger_efficiently_2022, dalleiger_federated_2025, hess_c-salt_2017}.

\begin{table*}[h!]
\centering
\renewcommand{\arraystretch}{1.2}
\begin{tabular}{>{\raggedright\arraybackslash}p{4cm}
                >{\raggedright\arraybackslash}p{1cm}
                p{10cm}}\hline
\textbf{Method} & \textbf{Finds rank?} & \textbf{Notes and reason for exclusion/inclusion} \\ \hline
ASSO \cite{miettinen_discrete_2008} & No & MDL4BMF, which was implemented, is based on ASSO but iterates K. \\
Delegation-Relegation, \cite{avellaneda_delegation-relegation_2024} & No & Exact approach limited to medium sized matrices (shows taking 3 hrs to solve on medium matrix, $\sim$3300 $\times$ 1600). Does not compare to PRIMP or other established BMF approaches. \\
Iteress \cite{belohlavek_factorizing_2019} & No & Does not compare to established BMF approaches, exact approach demonstrated on medium datasets only. Slow runtime in \citet{avellaneda_delegation-relegation_2024}. \\
ELBMF \cite{dalleiger_efficiently_2022} & No & Shows a very similar performance to PRIMP and is similar methodologically to PRIMP. Consider for future work. \\
MaxSAT under-cover, \cite{avellaneda_undercover_2022} & No & Solves a different problem, which does not allow for false positives. \\
Kovacs et al. \cite{kovacs_binary_2021} & No & Exact solution, not scalable. Shows solutions after 20 minutes (still not exact) on problem size maximum 105 $\times$ 105. \\
Bias-aware BMF, \cite{wan_bias_2022} & No & No indication of time for compute. Only solves problems for low K ($\leq$ 5). No comparison to PRIMP. \\
Message Passing, \cite{ravanbakhsh_boolean_2016} & No & No indication of compute time, only evaluate on small/medium matrices (1000 $\times$ 1000), and only compares to ASSO. \\
OrMachine, \cite{rukat_bayesian_2017} & No & Bayesian. Does not compare to established BMF methods. Evaluated on medium-small datasets. \\
MEBF \cite{wan_fast_2020} & Yes & We implemented MEBF for preliminary results and found it performed poorly, likely because no guidance is given on hyperparameter choice in related codebase. Also performed poorly in recent works, \cite{dalleiger_federated_2025, yang_boolean_2024}. \\
MDL4BMF \cite{miettinen_mdl4bmf_2014} & Yes & Implemented. Based on ASSO, one of the original BMF approaches consistently implemented across literature, \cite{hess_primping_2017, wan_bias_2022, dalleiger_efficiently_2022}. Does rank selection. \\
PANDA+ \cite{lucchese_unifying_2014} & Yes & Implemented. One of the few methods that also does rank selection, and used across benchmarks in literature \cite{hess_primping_2017, wan_bias_2022, makhalova_-below_2021}. \\
PRIMP \cite{hess_primping_2017} & Yes & Implemented. Consistently high-performing method across literature \cite{dalleiger_efficiently_2022, dalleiger_federated_2025, hess_c-salt_2017}. \\
\hline
\end{tabular}
\caption{Comparison of various BMF methods and reasons for exclusion/inclusion.}
\label{tab:bmf_comparison}
\end{table*}

\subsection{Illustrative example of RMP-1 cost}\label{sec:ill_eg}
The cost associated with a factor and a sample is given by $c_{i, \alpha} = \min\left ( \sum_{j| (i,j) \in E} \delta_{\alpha, j}, \sum_{j| (i,j) \notin E}  \delta_{\alpha,j} \right )$, where $\delta_{\alpha,j} = 1$ if factor $\alpha$ contains feature $j$.
When the true factors share features, the RMP-1 objective will favour disjoint factorisations even without explicit constraints due to the cost definition.
In Figure \ref{fig:ill_eg}, the lowest-cost solution contains three candidate factors despite the two true patterns existing in the candidate factors.
This illustrates that the RMP-1 formulation implicitly favours disjoint factorisation with higher rank.

However, consider that for the optimal feature set, $s_1$, only the first and third columns of the membership matrix differentiate the three groups, and these have the same pattern of membership ($L$) as all the other sets, where $K = 2$.
After identification of the best disjoint factors, if observations containing the same factor are grouped, then the true features corresponding to this grouping capture the underlying patterns $p_1, p_2$, and the unnecessary factor $\alpha_3$ can be dropped.
Algorithms 1 and 2 in the main text formalise this approach to uncover the true factorisation from the disjoint one.

\subsection{Candidate Factor Generation}\label{sec:cand_gen}
To generate candidate factors for our RMP, we perform hierarchical clustering across features,
based on their pairwise hamming distance, and cut the
hierarchical tree at several different levels. Each resultant
cluster from each level is taken as a candidate factor.
We also perform Leiden community detection at different resolutions (a hyperparameter), which initially constructs a K-nearest-neighbour graph, again based on hamming distance.
We take the union of clustered features as candidate factors to construct the RMP-w.

In the implemented code, candidate columns are generated by providing a target range of candidate columns for RMP-w, with the idea that more candidate columns will likely be more informative. 
The code will automatically increase or decrease the number of hierarchical cuts to meet this target, although there is an upper limit on the number of candidate columns (i.e all cuts of the hierarchical tree).
This is a hyperparameter that, if the model is underperforming, increasing this will likely improve results.
Results are shown with an upper limit of 10000 columns for simulated settings and 20000 for real data settings (constant over each setting), indicating relative robustness to the number of columns provided.

\subsection{Delayed Column Generation}\label{sec:col_gen}
The pricing problem (PP) for the restricted master problem is given by:

\begin{align}
    \min \sum_{i=1}^M t_i &- \sum_{j=1}^N \pi_j^*x_j - \gamma^* \\
    \text{s.t} \quad 
    t_i &= \min \left ( \sum_{j| (i,j) \in E} x_j, \sum_{j| (i,j) \notin E}  x_j \right ) \label{e1:pp_con} \\
     x_j &\in \{0, 1\}, \qquad \forall j \in N \notag \\
     t_i &\in \mathbb{R^+}, \qquad \forall i \in M \notag
\end{align}

In practice, to model linearly, constraint \ref{e1:pp_con} requires four constraints to implement, using a switch binary variable $s_i$ for each $M$.
Here $\pi_j^*$ is the value of the dual variable of constraint 6 (main text) at the optimal solution of the restricted master problem and $\gamma^*$ is the value of the dual of constraint 5 (main text).

\begin{figure}[t]
    \centering
    \includegraphics[width=\linewidth]{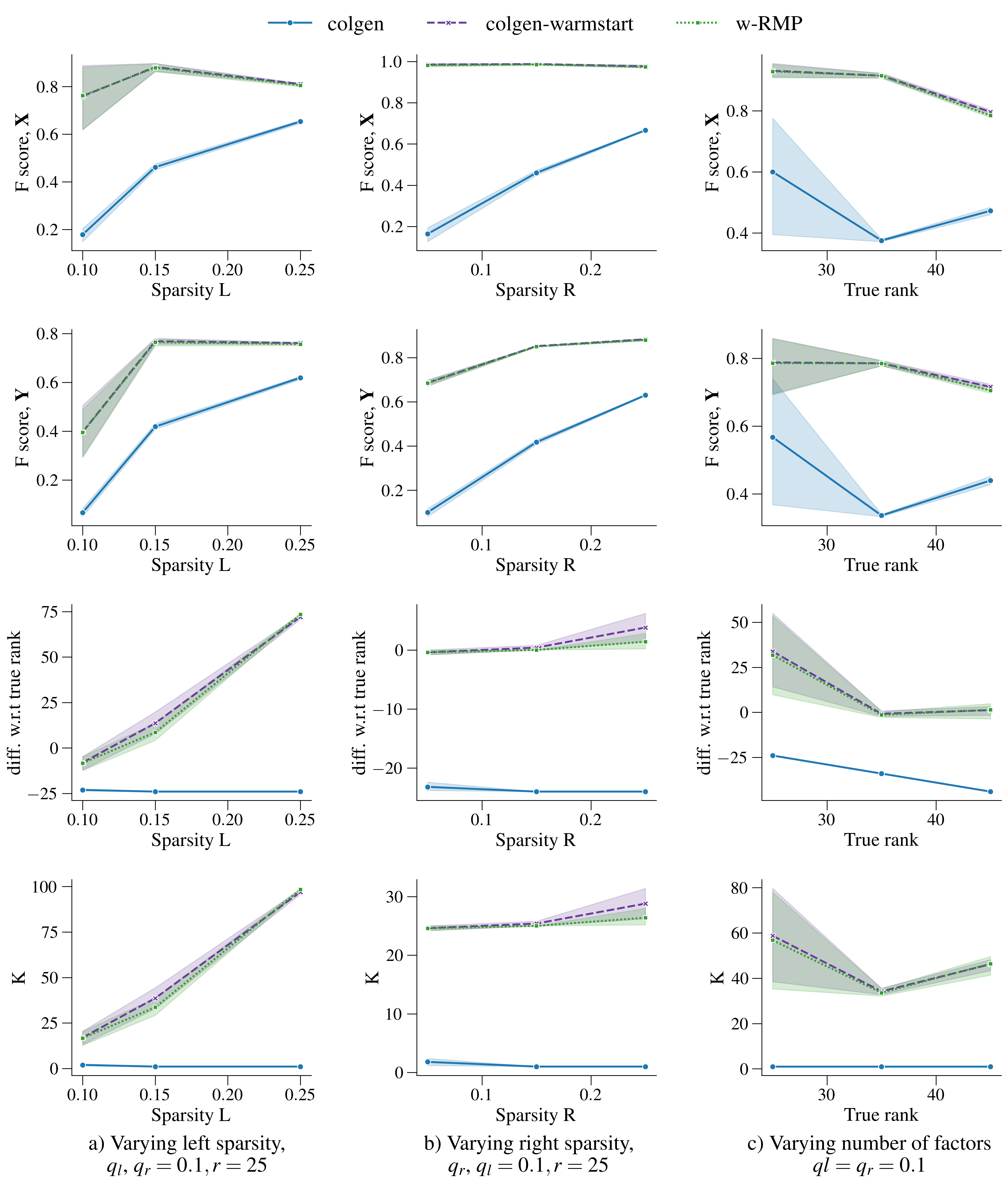}
    \caption{Comparison of delayed column generation used in tandem with bfact, simulations with $p^\pm = 0.1, M\times N= 1600 \times 400$. Each case is followed by the heuristic postprocessing with MDL cost, run only once for $K_{\min} = K_{\max} = 100$.} \label{fig:app_colgen}
\end{figure}

Here, the number of (variables, constraints) of the PP scales with $(M + N , \; M)$.
We found this to be slow.
We demonstrate using the PP with the RMP, both warm-started or not, also comparing to RMP-w alone in Figure \ref{fig:app_colgen}, where the delayed column generation process is capped at 30 minutes.
Given the RMP-w alone takes approximately 10 minutes, this demonstrates the PP does not offer a significant/timely advantage to the RMP, when warm-started appropriately.

\subsection{MDL cost}\label{sec:mdl}
The cost measure in PRIMP is given by the code-table cost:
\begin{align}
    f_{CT}(\mathbf{L}, \mathbf{R}, \mathbf{Y}) =& f_{CT}^D(\mathbf{L}, \mathbf{R}, \mathbf{Y}) + f_{CT}^{\cal M}(\mathbf{L}, \mathbf{R}, \mathbf{Y})  \label{eq:ft_cost} \\
    f_{CT}^D(\mathbf{L}, \mathbf{R}, \mathbf{Y}) =& -\sum_{k=1}^K |L_{\cdot k}|\log(p_k) - \sum_{j=1}^N |\eps_{\cdot j}|\log(p_{K+j}) \notag \\ 
     f_{CT}^{\cal M} (\mathbf{L}, \mathbf{R}, \mathbf{Y}) =&\sum_{k: |L_{\cdot k}| > 0} (R_{k \cdot}c - \log(p_k)) \notag \\ &+ \sum_{j: |\epsilon_{\cdot j}| > 0} (c_j - \log(p_{K+j})) \notag ,
\end{align}
where $\eps$ is the difference between $\mathbf{Y}$ and the reconstructed matrix, and the probabilities $p_k$ and $p_{K+j}$ refer to the usage of non-singleton profiles $R_{k\cdot}$ and singleton profiles $\{j\}$ (i.e profiles containing only a single feature).
These are given by:
\begin{align*}
    p_k = \frac{|L_{\cdot k}|}{|\mathbf{L}| + |\eps|}, \quad p_{K+j} = \frac{\eps_{\cdot j}}{|\mathbf{L}| + |\eps| }
\end{align*}
Further, $c: N\times 1$ is the vector of code lengths for each feature, given by, $c_j = - \log (|Y_j|/|\mathbf{Y}|)$.

\subsection{Limitations of MDL Code table cost} \label{sec:mdl_lim}
\begin{figure}[h!]
    \centering
    \includegraphics[width=0.5\linewidth]{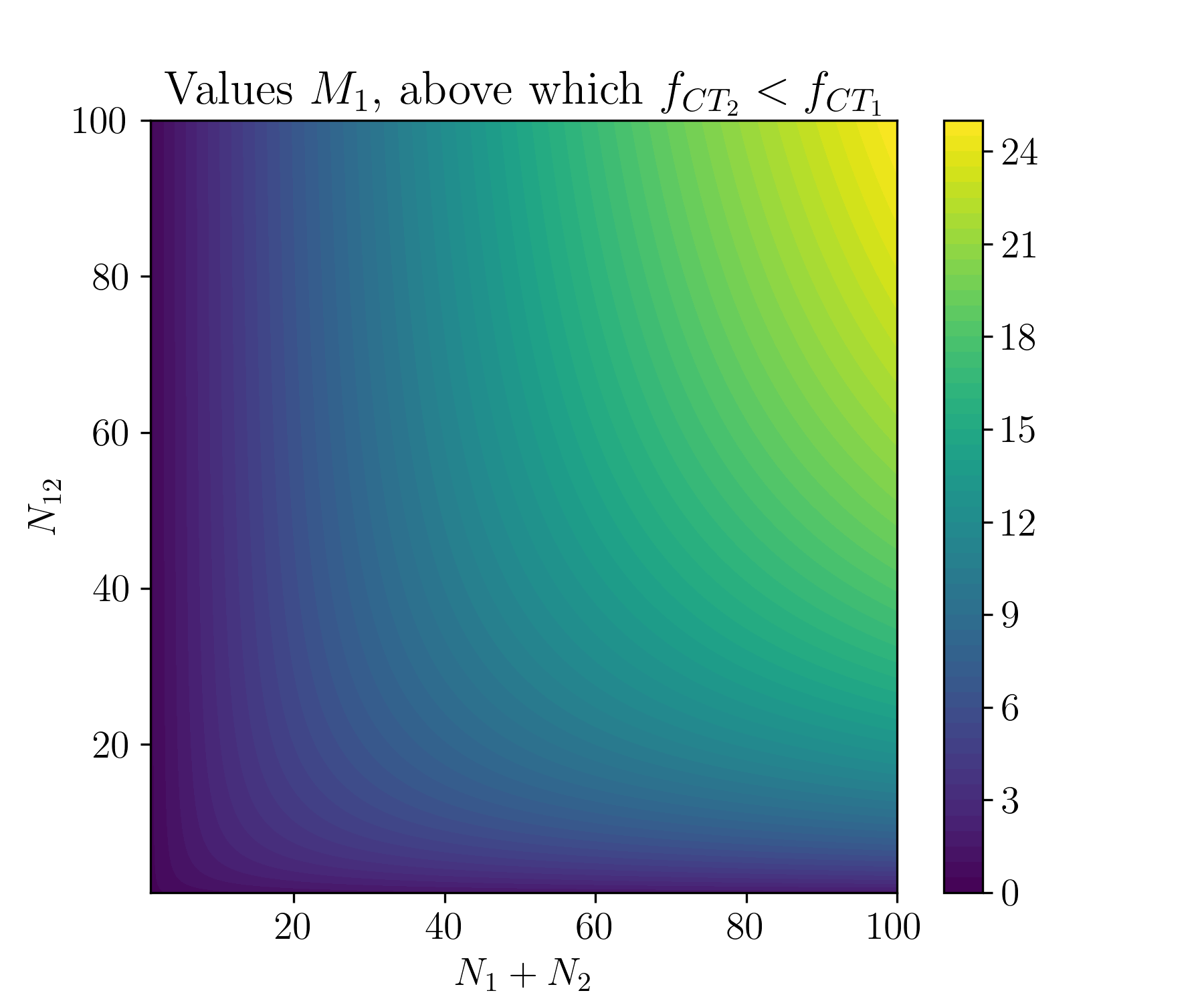}
    \caption{Demonstration of how higher rank representation is often favoured by MDL cost.}
    \label{fig:surf}
\end{figure}

The MDL Code table cost also favours disjoint representations, leading to sparsity in the left and right decomposed matrices. To demonstrate, consider two true underlying factors- let each factor have $N_1, N_2$ unique features and share $N_{12}$ features. Also, let the number of observations that contain only one of each factor be $M_1, M_2$, and the number that includes both be $M_{12}$. For simplicity, let $M_1 = M_2 = M_{12}$. We consider two scenarios here - the first, where we recapitulate the true factors, and the second, where we consider three factors, each corresponding to the unique features of true factors 1 and 2, and a third factor for their shared features. Here, $\bfeps = 0$ as both these decompositions exactly reconstruct the input data $\mathbf{Y}$. Then the MDL cost in the first instance is:

{\scriptsize
\begin{align}
    f_{CT_1} =& - (M_1 + M_{12})\log\frac{M_1 + M_{12}}{M'}  - (M_2 + M_{12})\log\frac{M_2 + M_{12}}{M'} \notag \\
    &- N_1\log\frac{N_1}{N'} -  N_2\log\frac{N_2}{N'} -  2N_{12}\log\frac{N_{12}}{N'} \notag \\
    =& -4M_1\log2/3 - N_1\log\frac{N_1}{N'} -  N_2\log\frac{N_2}{N'} -  2N_{12}\log\frac{N_{12}}{N'}  \\
    f_{CT_2} =& - M_1\log\frac{M_1}{M'}  - M_2\log\frac{M_2}{M'} - M'\log\frac{M'}{M'}- N_1\log\frac{N_1}{N'} \notag \\
    &-  N_2\log\frac{N_2}{N'} -  N_{12}\log\frac{N_{12}}{N'} \notag \\
    =& -2M_1 \log1/3 - N_1\log\frac{N_1}{N'} -  N_2\log\frac{N_2}{N'} -  N_{12}\log\frac{N_{12}}{N'} 
\end{align}
}
Where $N' = N_1 + N_2 + N_{12}$ and $M' = M_1 + M_2 + M_{12} = 3M_1$.
\medbreak
Taking the difference between the higher rank decomposition (i.e with three factors), and the lower rank, we get:
{\footnotesize
\begin{align}
    f_{CT_2} - f_{CT_1} &= -2M_1 \log1/3 + 4M_1\log2/3 + N_{12}\log\frac{N_{12}}{N'} \notag \\
    &= -2M_1 \log1/4+ N_{12}\log\frac{N_{12}}{N'} 
\end{align}
}
Plotting the surface; $M_1 = \frac{N_{12}}{2\log1/4}\log\frac{N_{12}}{N'}$, we get the values for $M_1$ above which the higher rank matrix has a lower MDL cost than the lower rank matrix, Figure \ref{fig:surf}.
\medbreak
The intuition behind this is that higher-rank representations are sparser, as the repeated features are not included in both factors in the right matrix $\mathbf{\hat{R}}$, and the slight increase in density in the left matrix (for associating an observation with another factor) $\mathbf{\hat{L}}$ is not enough to offset the shared features.
The same issue applies to the MDL regularisation given by $|\mathbf{\hat{L}}|$ and  $|\mathbf{\hat{R}}|$.
Although the typed XOR MDL cost used in \cite{miettinen_mdl4bmf_2014} includes an explicit reference estimated rank, $K$, we empirically found it also suffers from the same issue.

\subsection{Evaluation metric details}\label{sec:eval_met}
The F1 score is the harmonic mean of the precision and recall of the ones in the matrix $\mathbf{\hat{X}} = \mathbf{\hat{L}}\mathbf{\hat{R}}$ compared with either the signal matrix $\mathbf{X}$ (when known, in simulated settings), or the observed data matrix $\mathbf{Y}$.
This is given by:
\begin{align*}
    p_X &= \frac{\sum_{ij} X_{ij}\hat{X}_{ij}}{\sum_{ij} \hat X_{ij}} \qquad p_Y = \frac{\sum_{ij} Y_{ij}\hat{X}_{ij}}{\sum_{ij} \hat X_{ij}}\\
    r_X &= \frac{\sum_{ij} X_{ij}\hat{X}_{ij}}{\sum_{ij} X_{ij}} \qquad r_Y = \frac{\sum_{ij} Y_{ij}\hat{X}_{ij}}{\sum_{ij} Y_{ij}}\\
    F_{1,X} &= \frac{2 \times p_X \times r_X}{p_X + r_X} \qquad F_{1,Y} = \frac{2 \times p_Y \times r_Y}{p_Y + r_Y} 
\end{align*}

\subsection{Method Implementation Details}\label{sec:meth_imp}
\textbf{PRIMP}
We implemented PRIMP to run for $50000$ steps, following what the authors did in \citet{hess_primping_2017}, and used a $\Delta_k = 5$, from $5$ to $100$.
PRIMP was implemented on simulations with access to 6 CPUs, and 1 NVIDIA GPU (of varying specifications, mostly Quadro RTX 8000 or  P100 SXM2). 
\medbreak
For the real data, PRIMP was run for $50000$ steps, used a $\Delta_k = 10$, from $10$ to $100$ (the method stops at $\max_K + \Delta_k$).
Again it was run on machines with access to 6 CPUs and 1 NVIDIA GPU.

\textbf{PANDA}
The PANDA documentation provides little guidance on what hyperparameters to use. We tested all and used the best combination, which was a frequency strategy, and a type 1 cost.
It was run with a maximum $K$ of 100 for both simulated and real scenarios.

\textbf{MDL4BMF}
MDL4BMF takes longer than the other methods to run, we implemented simulations with 12 CPUs, double of other methods.
We implement it with hyperparameters following the README example given in the code- with 10 threshold parameters and all error measures.
For real data, we ran each dataset with access to 24CPUs, terminating if the model had not completed within 2 days. 

\textbf{bfact}
For matrices lower than a certain size, $M \times N < 5e6$, we transpose the matrix if $N < M$.
We use a $\Delta_k = 10$, from $10$ to $100$.
For the reconstruction procedure, we used $f = 0.997$ for simulated data. 
As a rule of thumb, found $f = \min(1 - 1/\min(M, N), 0.999)$ truncated at 3 decimals works well, which we use for the real data matrices.





\subsection{Extended results}\label{sec:ext_res}
\begin{figure}[h!]
    \centering
    \includegraphics[width=0.5\linewidth, trim=0 0 8pt 15pt, clip]{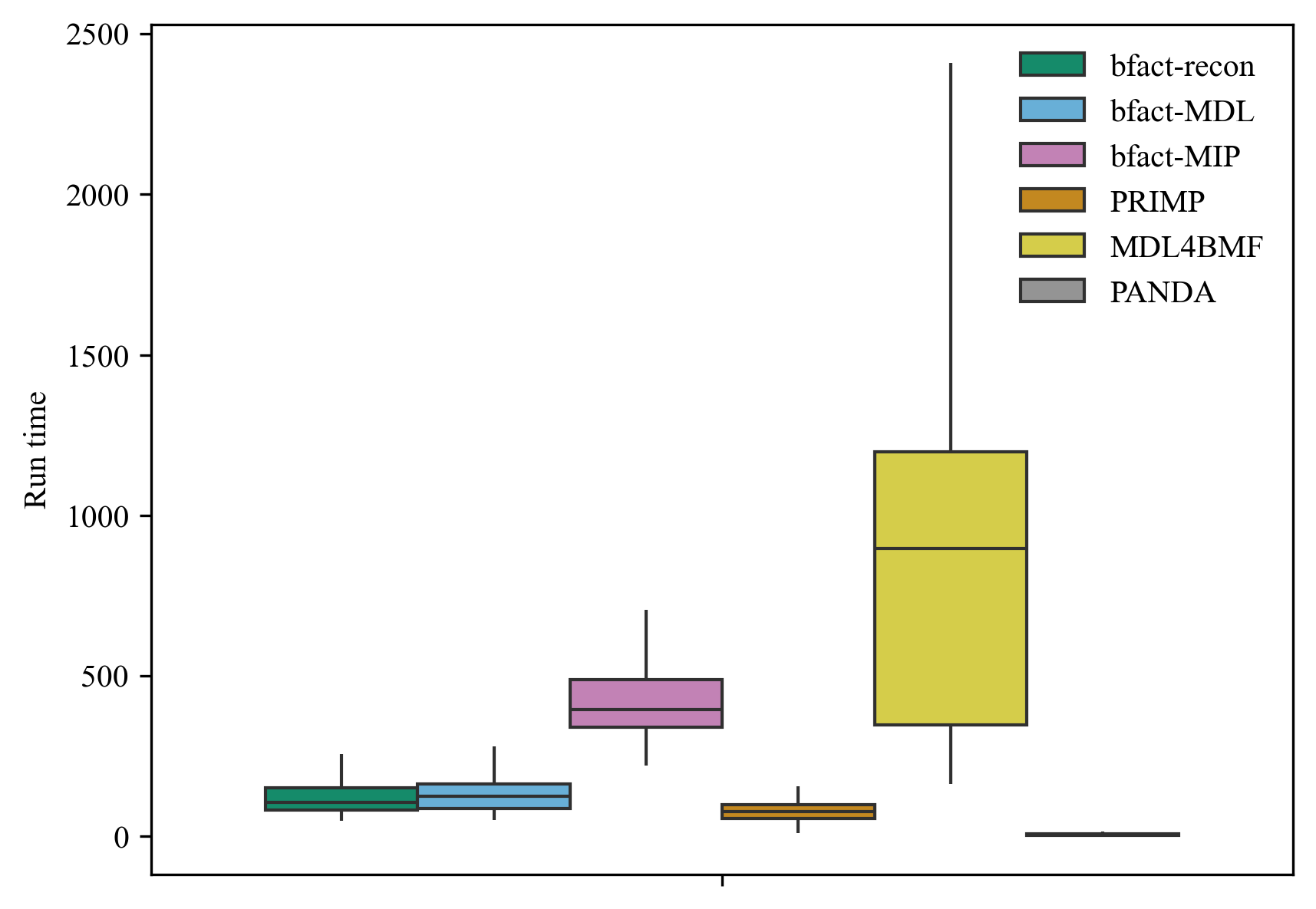}
    \caption{Run time across simulations}
    \label{fig:sim_time}
\end{figure}

\begin{figure*}[h!]
    \centering
    \includegraphics[width=\linewidth]{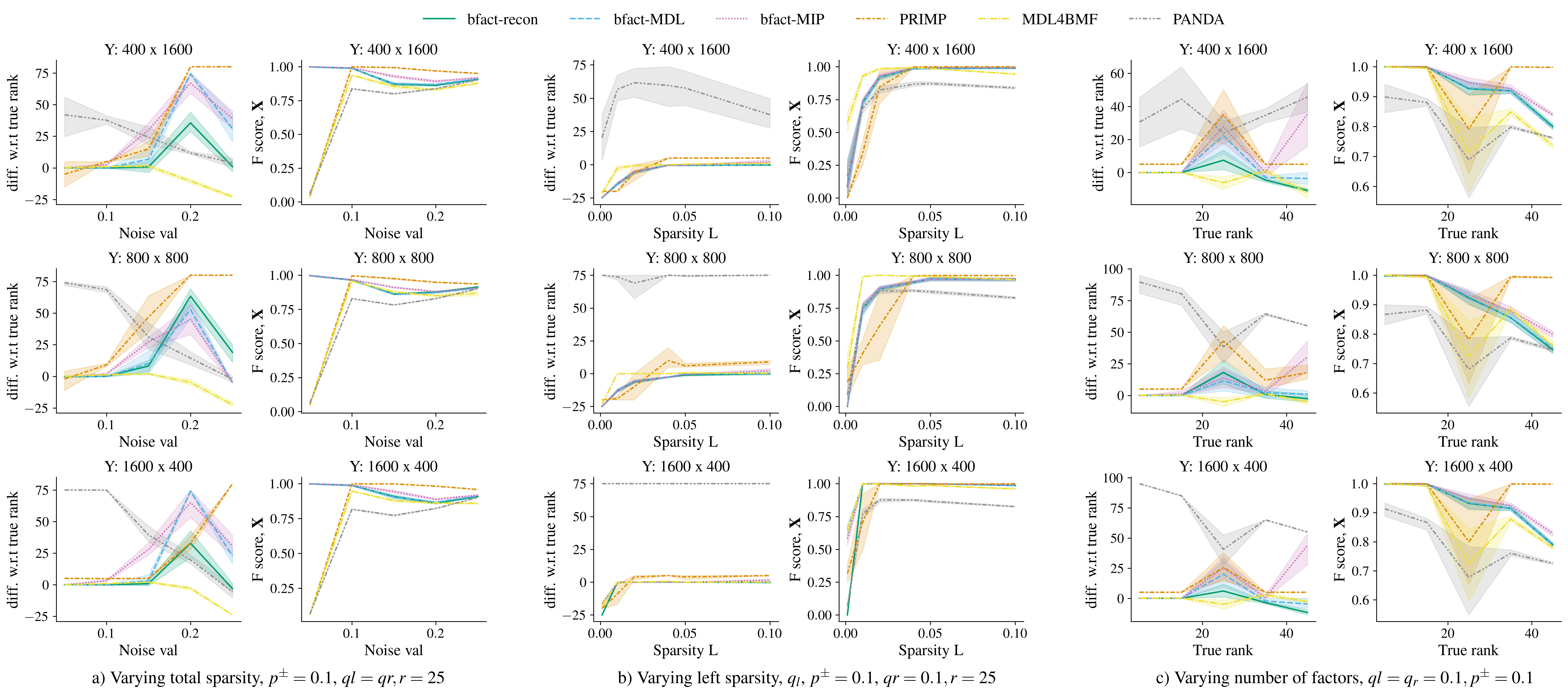}
    \caption{Further simulation results, part A} \label{fig:app_sim_results_A}
\end{figure*}

\begin{figure*}[h!]
    \centering
    \includegraphics[width=\linewidth]{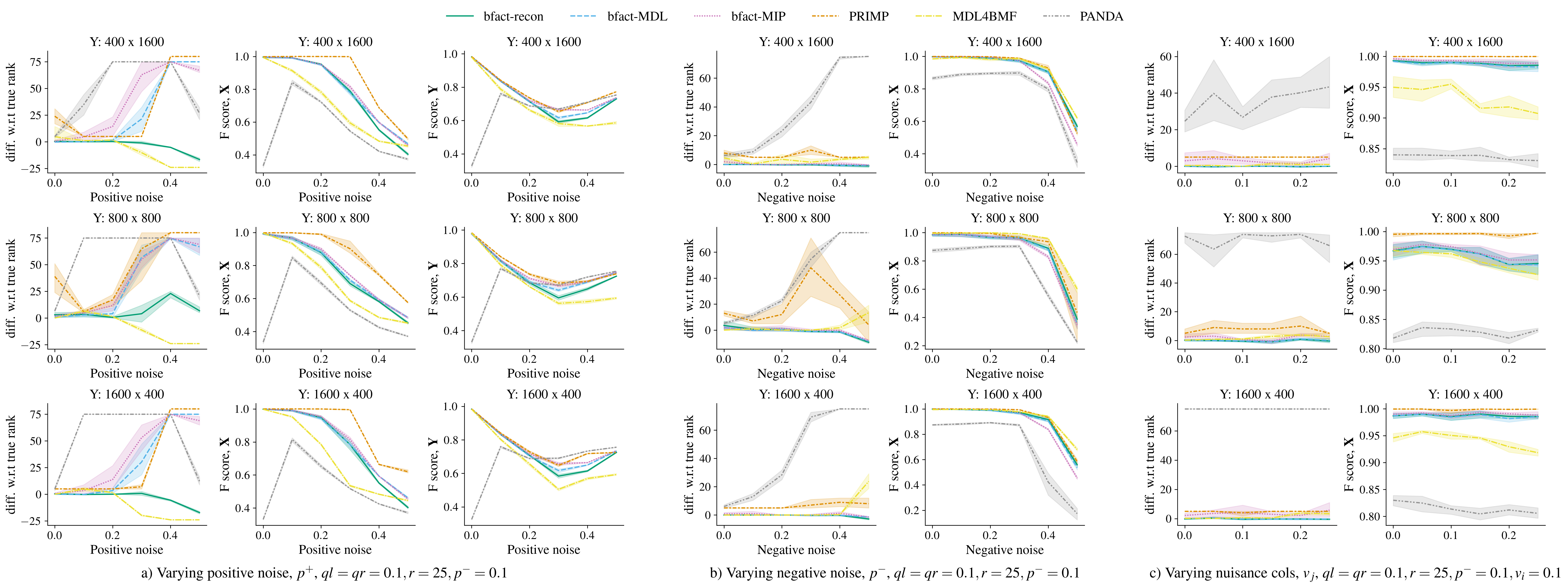}
\caption{Further simulation results, part B, for a) we also include the F-score on the data matrix, $\mathbf{Y}$ to show that a higher score here does not necessarily translate to a higher score for $\mathbf{X}$, due to overfitting to noise. } \label{fig:app_sim_results_B}
\end{figure*}

Figure \ref{fig:sim_time} details the runtime of different methods across simulations, while Figure \ref{fig:app_sim_results_A} and \ref{fig:app_sim_results_B} detail resuts of methods across different simulation settings.

\end{document}